\documentclass[lettersize,journal]{IEEEtran}

\usepackage{amsmath,amsfonts}
\usepackage[ruled,linesnumbered]{algorithm2e}
\usepackage{array}
\usepackage{multirow}
\usepackage[caption=false,font=normalsize,labelfont=sf,textfont=sf]{subfig}
\usepackage{textcomp}
\usepackage{stfloats}
\usepackage{url}
\usepackage{verbatim}
\usepackage{graphicx}
\usepackage{cite}
\usepackage{hyperref}
\begin{document}
\pagestyle{empty}

\title{Model-Free Privacy Preserving Power Flow Analysis in Distribution Networks}

\author{Dong~Liu,
        Juan S. Giraldo,~\IEEEmembership{Senior Member,~IEEE,}
        Peter~Palensky,~\IEEEmembership{Senior Member,~IEEE,}
        and Pedro~P.~Vergara,~\IEEEmembership{Senior Member,~IEEE}
\thanks{Dong Liu, Pedro P. Vergara (corresponding author) and Peter Palensky are with the Intelligent Electrical Power Grids (IEPG) Group, Delft University of Technology, Delft 2628CD, The Netherlands (e-mail: p.p.vergarabarrios@tudelft.nl). 

Juan S. Giraldo is with the Energy Transition Studies Group, Netherlands Organisation for Applied Scientific Research.
}}




\maketitle
\thispagestyle{empty}

\begin{abstract}
Model-free power flow calculation, driven by the rise of smart meter (SM) data and the lack of network topology, often relies on artificial intelligence neural networks (ANNs). However, training ANNs require vast amounts of SM data, posing privacy risks for households in distribution networks. To ensure customers' privacy during the SM data gathering and online sharing, we introduce a privacy preserving PF calculation framework, composed of two local strategies: a local randomisation strategy (LRS) and a local zero-knowledge proof (ZKP)-based data collection strategy. First, the LRS is used to achieve irreversible transformation and robust privacy protection for active and reactive power data, thereby ensuring that personal data remains confidential. Subsequently, the ZKP-based data collecting strategy is adopted to securely gather the training dataset for the ANN, enabling SMs to interact with the distribution system operator without revealing the actual voltage magnitude. Moreover, to mitigate the accuracy loss induced by the seasonal variations in load profiles, an incremental learning strategy is incorporated into the online application. The results across three datasets with varying measurement errors demonstrate that the proposed framework efficiently collects one month of SM data within one hour. Furthermore, it robustly maintains mean errors of 0.005 p.u. and 0.014 p.u. under multiple measurement errors and seasonal variations in load profiles, respectively.

\end{abstract}


\section{Introduction}

\IEEEPARstart{P}{ower} flow (PF) calculation is an essential tool for the operation and planning of distribution networks (DNs). Traditional power flow calculation requires the topology and parameters of the network topology \cite{guo2021data}. However, this information is not normally available, especially in low-voltage distribution networks (LVDNs). To address this issue, model-free PF calculation approaches have been proposed based on deep learning models such as artificial neural networks (ANNs) \cite{10176302}. By training on extensive datasets that encompass a wide range of operational scenarios, ANNs achieve high estimation accuracy for voltage magnitude, angle and missed measurements \cite{hu2020physics}. The training of an ANN requires a large amount of training data, including active power, reactive power, voltage magnitude and phase angle. However, the process of collecting the smart meter (SM) data might leak privacy information (e.g., appliance brands) and customer behaviour (e.g., working from home time) 
\cite{9548947}. Privacy concerns hinder measurement sharing in DNs and the development of model-free PF approaches.

Traditional encryption algorithms, like symmetric encryption, can safeguard SM data during transmission \cite{9462936,10400180}. However, decrypted data is essential for complex calculations like optimization problems and PF calculations. This decryption exposes household privacy to potential cyber threats at the control centre. Existing privacy-preserving approaches for measurement sharing in DNs could be roughly divided into equipment-based and algorithm-based categories. Equipment-based approaches, such as using house energy storage systems (ESSs) to alter shared data, can introduce variations that obscure real distributions of SM data \cite{zheng2021decentralized}. Nevertheless, the high installation cost of ESS constrains its applicability for privacy protection and is mainly used for absorbing distributed energies. Moreover, new algorithms or indicators are needed to trade off data privacy and utility, enabling the altered data to be compatible with existing algorithms at the distribution system operator (DSO). 
\par\IEEEpubidadjcol

Algorithm-based approaches integrate homomorphic encryption (HE), data aggregation and randomization. A fully HE approach supports homomorphic addition and multiplication, i.e., the decrypted results obtained based on encrypted data is the same as the results obtained based on unencrypted data. This characteristic of HE enables SMs to share the encrypted data with the DSO to participate in the calculation and receive the encrypted results. A privacy protection strategy based on partial HE was proposed in \cite{wu2021privacy} to ensure the privacy of the updated parameters during alternating optimization in optimal power flow (OPF) calculation. HE was combined with an aggregation strategy to ensure household privacy in \cite{busom2016efficient}, which does not result an intense calculation burden for SMs. To ensure privacy in energy management systems, distributed optimization models with HE were proposed and their convergence was verified in \cite{yuan2023fully}. However, the traditional HE algorithm is inefficient and only supports homomorphic addition and multiplication, which does not guarantee the accuracy and convergence of the classical PF model.

Noise disturbance is a straightforward and effective strategy for data randomization, making them widely utilized in data privacy protection. Differential Privacy (DP), a leading noise disturbance strategy, is commonly used to enhance the privacy of data \cite{huang2022dpwgan}. A Laplace noise disturbance strategy is employed to enhance the privacy level and accuracy of distributed dispatch problems \cite{sun2023privacy}. In \cite{9146187}, a local randomization approach was introduced to independently ensure the privacy of users and noise disturbance was employed to enhance the privacy level. Two low-cost and lightweight DP strategies were introduced to reduce the expense of ESS-based privacy-preserving approaches in \cite{7501552}. To trade off the privacy level and accuracy loss induced by injected noise, the results in \cite{yang2017differential} show that the OPF model with DP strategy should be further remodelled. Instead of injecting noise into network parameters and inputs, OPF variables are modelled as a function of random noises, guaranteeing the feasibility and privacy level of the model \cite{dvorkin2020differentially}. Nevertheless, DP is applied to ensure privacy in OPF and distributed optimization problems in the above papers, and its application in power flow calculations is quite rare.

DP strategies can also be integrated into ANNs to ensure the privacy of training datasets, e.g., backwards gradient with noises \cite{9993791}. Nevertheless, several limitations hinder their application for PF calculation and are summarized as 1) The generalization of ANN is inherently limited, posing a challenge in achieving a similar accuracy when an ANN is trained on original data and data with noise \cite{9743327}. 2) Adding noise (e.g., Laplace or Gaussian noise) to SM data might cause two different data points to be the same value, affecting the accuracy of PF calculation. Ensuring consistent monotonicity before and after data encryption is essential to maintain the accuracy of ANN-based PF calculation. 3) DP requirements on the similarity between adjacent datasets, stipulating that they should differ by only a single sample. While OPF approaches address this criterion by focusing on critical loads, ensuring the privacy of each SM data point in PF calculation is imperative, not just for critical loads \cite{dobbe2020local}. In summary, the privacy-preserving strategy for model-free PF calculation should trade off privacy level, data utility and accuracy.

Zero-knowledge proof (ZKP) is widely used in scenarios where privacy and confidentiality are paramount, such as authentication protocols and privacy-preserving blockchain technologies \cite{lavin2024survey}. ZKP enables the verification of transactions without revealing transaction details, enhancing privacy and scalability in the blockchain. A ZKP-based energy trading approach integrates the ZKP into the blockchain to ensure fair operation without cheating participants \cite{hou2022privacy}. To protect privacy in the hybrid energy dispatch, ZKP was adopted to ensure the accuracy of the profit variation, aiming to fairly anER economically distribute relocated profit in \cite{zhou2023incentivized}.
A privacy-preserving strategy was proposed based on blockchain and ZKP in \cite{gabay2020privacy} for safely authentication the connected vehicle in a distributed environment other than central authority.  Besides, variants of ZKP extend its usability by eliminating the need for interaction between the prover and verifier, such as Zero-knowledge succinct non-interactive arguments of knowledge. Nevertheless, the application of ZKP in privacy-preserving and dataset-sharing in DNs is still in its early stages.

On the other hand, the statistical distribution of SM data varies across seasons due to differences in the use of electrical devices and energy consumption behaviours, such as the increased use of heating pumps during winter and the charging patterns of ESSs. These seasonal variations impact the accuracy of the ANN-based PF analysis. Re-training the ANN with updated data to maintain accuracy introduces substantial computational overhead, and the accuracy of the updated ANN on old datasets may significantly decrease. Incremental Learning (IL) strategies are proposed to mitigate catastrophic forgetting during ANN updating by preserving knowledge acquired from old data while learning new information from the latest datasets. IL strategies can be broadly categorized into three clusters: Architecture-based, Regularization-based, and Rehearsal-based  \cite{van2022three,zenke2017continual}. Nevertheless, IL strategies have been extensively applied in classification tasks, and their application in enhancing the generalization of ANN-based PF approaches in DNs remains relatively underexplored.

To address this research gap, this paper introduces a privacy-preserving PF calculation framework, composed of two local strategies: a local randomisation strategy (LRS) and a local ZKP-based data collection strategy. These are novelty integrated into ANN-based PF analysis to guarantee the security of the shared SM data. Moreover, an IL strategy is employed to maintain the accuracy of ANN-based PF results under datasets with multiple season patterns. The main contributions of this paper can be summarized as follows:

\begin{itemize}
    \item When sharing power data with DSOs aiming to perform PF calculations, a local strategy is proposed to randomize reactive and active power, utilizing modified Sigmoid and random functions. The proposed strategy does not rely on third parties and neither on DP frameworks.
    \item A ZKP-based data collection strategy is proposed to collect the training voltage dataset, without inducing privacy leakage. This strategy enables SMs to individually confirm the utility and correctness of the voltage dataset without directly sharing actual voltage values with DSO.    
    \item To conduct PF analysis using privacy-transformed power data, a data-driven approach is essential—one capable of accurately mapping the transformed SM data. Thus,  a model-free PF analysis scheme is introduced to handle the complexities introduced by data transformations.
    \item To address accuracy loss induced by seasonal load variations, an IL strategy is used to periodically update the ANN, triggered by a Wasserstein distance (WD)-based indicator. This strategy ensures PF calculation accuracy by adapting to temporal variations.

\end{itemize}

\section{Preliminaries}
\label{section2}

\subsection{Model-free Power Flow Model}
Power flow calculation aims to calculate the unknown variables of each bus in the distribution networks, including reactive power, active power, voltage magnitude and phase angle, depending on the type of bus. The relation among these four variables is formulated as expressions \eqref{eq1} and \eqref{eq2}. To calculate the PF in a DN with $N$ buses and no distributed generators, there are $2N$ unknown variables—typically the voltage magnitudes and phase angles at each bus. To determine the unknown values, it is necessary to formulate and solve a corresponding set of $2N$ nonlinear equations that represent the power balance at each bus.
\vspace{-0.0cm}
\begin{align}
\label{eq1}
&P_n = V_n \sum_{m=1}^{N} V_m (G_{nm} \cos\theta_{nm} + B_{nm} \sin\theta_{nm})\\
\label{eq2}
&Q_n = V_n \sum_{m=1}^{N} V_m (G_{nm} \sin\theta_{nm} - B_{nm} \cos\theta_{nm})
\end{align}

where $G_{nm}$ and $B_{nm}$ are the conductance and susceptance of the connection lines between nodes $n$ and $m$. $P_n, Q_n$ and $V_n$ are the active power, reactive power and voltage magnitude at node $n$, respectively. $\theta_{nm} = \theta_n - \theta_m$ is the phase angle difference between nodes $n$ and $m$.

The Newton-Raphson method is a widely used iterative technique for solving PF equations. The topology of DNs is required in the Newton-Raphson method, which includes the connection lines, the length of cables, etc. However, this information is not normally available in DNs \cite{liu2025topology}. The model-free PF approach was proposed to address this issue, which aims to grasp the nonlinear relationship between unknown and known variables from historical measurements \cite{lin2024powerflownet}.

By leveraging the nonlinear mapping capabilities of ANN, voltage magnitude and phase angle could be estimated from active and reactive power. According to \eqref{eq1} and \eqref{eq2}, the inputs to the ANN could be the active power $P_n$ and reactive power $Q_n$ at each node. The outputs are the voltage magnitude $V_n$ and phase angle $\theta_n$. The estimation of voltage magnitude and angle are formulated as \eqref{eq3} and \eqref{eq4}.
\begin{align}
\label{eq3}
&V = F_{\psi1}(P_1, P_2, \ldots, P_N, Q_1, Q_2, \ldots, Q_N)\\
\label{eq4}
&\theta = F_{\psi2}(P_1, P_2, \ldots, P_N, Q_1, Q_2, \ldots, Q_N)\\
\label{eq5}
&[V_1,...,V_N, \theta_i,...,\theta_N] = F_{\psi3}(P_1, \ldots, P_N, Q_1, \ldots, Q_N)
\end{align}
where $F_{\psi1}$, $F_{\psi2}$ and $F_{\psi3}$ are the ANNs mapping the given SM data to voltage magnitude, angle and both, respectively.

To train an ANN-based PF model, a loss function commonly used is the mean squared error (MSE) $\mathcal{L}_\text{MSE}$ between the estimated and actual values, which are formulated as follows:
\begin{align}
\label{eq6}
&\mathcal{L}_\text{MSE} =\frac{1}{N} \sum_{n=1}^{N} \left[(V_n - \widetilde{V}_n)^2 + (\theta_n - \widetilde{\theta}_n)^2\right]
\end{align}
where $\widetilde{V}_n$ and $\widetilde{\theta}_n$ are the estimated voltage magnitude and angle from ANNs.
The ANN-based approach shows several advantages over traditional methods, including faster computation times, robustness to the error in SM data, and the ability to model complex, nonlinear relationships for DNs with unavailable topology \cite{hu2020physics}.
\subsection{Zero Knowledge Proof}
ZKP is a cryptographic protocol that allows one participant (named \textit{prover}) to convince another participant (named \textit{verifier}) that a given statement is true without revealing private information beyond the fact that the statement is true. The statement could be data values, a private calculation, a relationship, etc. A ZKP strategy must satisfy three fundamental properties: completeness, soundness, and zero knowledge. Completeness ensures that if the statement is true and the prover follows the protocol, the verifier will definitely trust the statement. Soundness guarantees that if the statement is false, no participant can convince the verifier, i.e., reject the statement. Zero-knowledge ensures that the verifier learns nothing except that the statement is true; specifically, the verifier gains no knowledge that could help reconstruct the prover secret.

Pedersen Commitment (PC) is a typical cryptographic commitment scheme, which is employed in privacy-preserving protocols and ZKP applications to enhance privacy and security. In ZKP scheme, PC allows the prover to commit to a value while keeping it hidden and later reveal it without altering the commitment. PC scheme is developed based on discrete logarithms to ensure that it is infeasible to change the committed value once it has been set, thus providing binding and hiding properties. Specifically, in ZKP schemes for range or equality proofs, a prover adopts PC to commit to value $x_0$ and then construct proofs demonstrating knowledge of these values or properties about them (i.e., proving that a committed value falls within a pre-set range or is the pre-set values) without revealing the actual values. PC based on the elliptic curve is formulated as follows:
\begin{align}
\label{eq0}
&C_{g,h}(x_0, r) = g^{x_0} \cdot h^r
\end{align}
where $g$ and $h$ are randomly sampled from the pre-set elliptic curve, respectively. $r$ is a random value within $[1, p]$, aiming to blind the true value. $p$ is a large prime number. The general process of PC-based ZKP is described as:
\begin{enumerate}
    \item Generate parameters $p$, $g$ and $h$.
    \item Prover generates an random value  $r$, and then create and send the commit $C_{g,h}(x_0,r)$.
    \item Verifier generates a challenge $b$ and sends it to the prover.
    \item Prover blinds the random value $r$ using (\ref{eq01}) and function $mod\{\cdot\}$\footnote{\textit{mod} \{$\cdot$\} is a function that returns the remainder of a division between two numbers.}, and the obtained $s$ is send to the verifier.
    \begin{equation}
        \label{eq01}
        s = mod \{r + x_0 \cdot (1-b), p\}
    \end{equation}
    \item Verifier checks the result using the private verification function $f^*$.
    \begin{equation}
        f^*(C_{g,h}(s,r), p, g, h, b, s) =  \begin{cases}
            1, & \text{if TRUE}\\
            0, & \text{if FALSE}
        \end{cases}
    \end{equation}
\end{enumerate}

Note that PC is homomorphic. This feature facilitates efficient ZKP constructions. This homomorphism enables the verifier to conduct addition directly on commitments, expressed as \eqref{eq02}. The formulation and detail of the ZKP process can be found in \cite{zhou2023incentivized}.
\begin{align}
\label{eq02}
&C_{g,h}(s1, r1) + C_{g,h}(s2, r2) = C_{g,h}(s1+s2, r1+r2)
\end{align}

\subsection{Incremental Learning}
IL strategies aim to mitigate catastrophic forgetting during ANN updating by preserving knowledge acquired from previous datasets while learning new information from the latest datasets. Compared to traditional class incremental learning, the IL strategy for PF analysis must handle complex ANN architectures due to high-dimensional output vectors. Architecture-based IL strategy is a widely used approach to preserving knowledge in continual learning scenarios. Within this category, freezing the weights of specific layers or reducing the learning rate are common and effective techniques \cite{ zenke2017continual}, as illustrated in (\ref{eq18}) and (\ref{eq19}).
\vspace{-0.0cm}
\begin{align}
\label{eq17}
& \mathcal{\psi}^{k-1} = [W^{k-1}, B^{k-1}] \\
\label{eq18}
& \mathcal{\psi}_{\textbf{\textit{l}}}^k = \mathcal{\psi}_{\textbf{\textit{l}} }^{k-1}\\
\label{eq19}
& \eta^k = \gamma . \eta^{k-1} 
\end{align}
where $\boldsymbol{l}$ denotes the index set of frozen layers. $\eta$ is the learning rate. $\psi^{k-1}$ and $\eta^{k-1}$ are the parameters and learning rate obtained at the end of the $(k-1)$-th update, respectively. $\gamma$ is a positive constant less than 1.

\begin{figure}
\centering
\includegraphics[width=0.48\textwidth,height=0.33\textwidth]{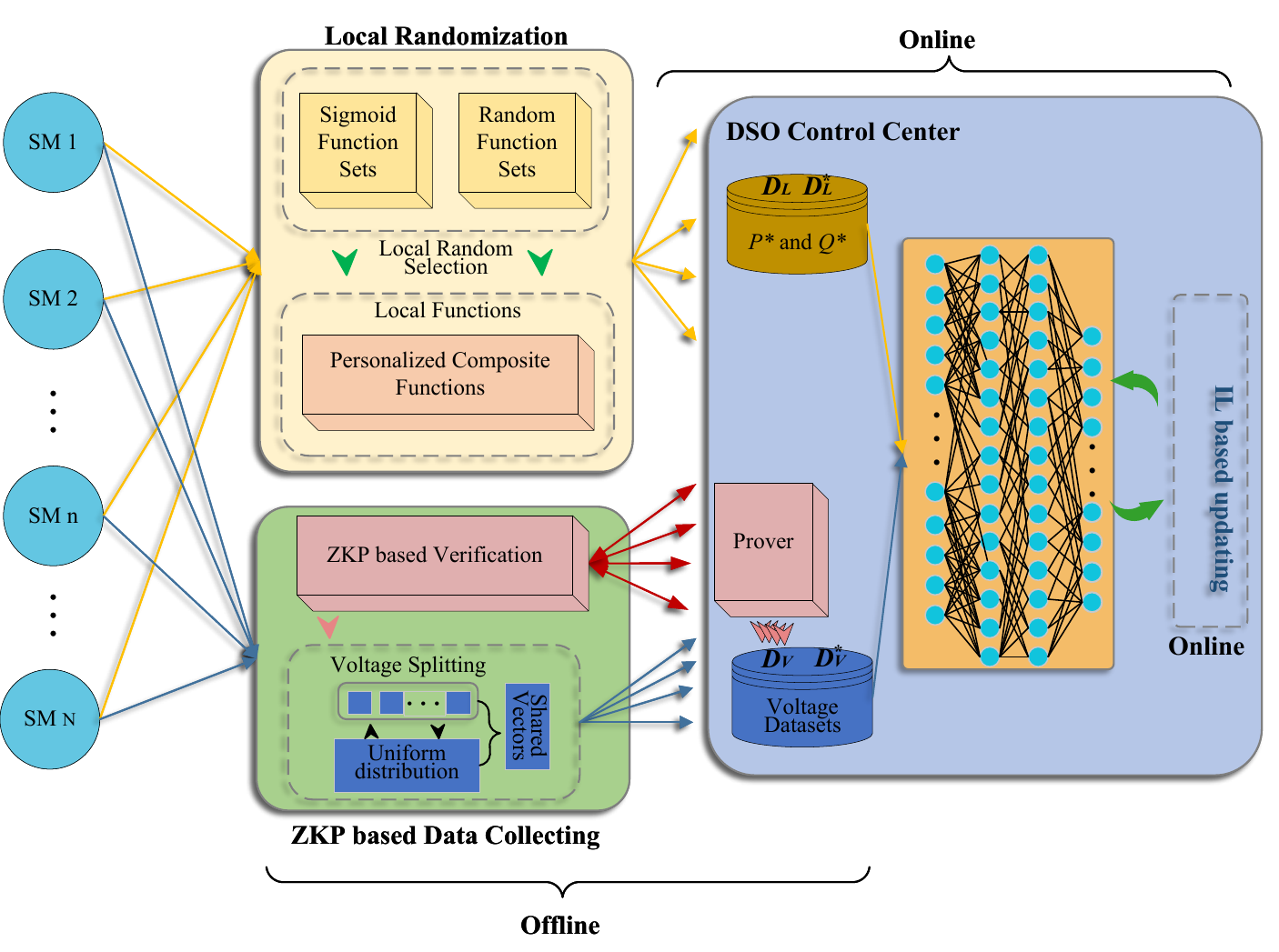}
\vspace{-0.5cm}
\caption{Framework of ANN-based PF calculation with local privacy-preserving strategy: orange, blue, and red lines represent load transfer, voltage transfer, and the communication channel, respectively.}
\vspace{-0.3cm}
\label{Fig.1}
\end{figure}

\section{Privacy preserving Power Flow Analysis}
\label{section3}
The proposed model-free power flow calculation framework with local privacy-preserving is illustrated in Fig \ref{Fig.1}.  We assume that bi-communication channels between SM and DSO exist for data sharing. In the offline phase, randomized power and voltage magnitude profiles are collected by DSO, which serves as the training dataset for the ANN. The privacy of the two datasets is ensured by the proposed \textit{Local Randomization Strategy} and the \textit{ZKP-based data collection strategy}, respectively. In the online phase, only randomized power data are shared with the DSO for estimating the voltage magnitude via an ANN. Moreover, an architecture-based IL strategy is employed to update the ANN model upon receiving datasets corresponding to different seasonal variations.

\subsection{Local Randomization Strategy}
\label{3-1}
The local randomization strategy for the active and reactive SM data must ensure the monotonicity, irreversibility and non-negativity of the data to maintain privacy levels and structural accuracy. Motivated by the randomization strategy in DP and the principle that addition operations are irreversible, we propose to locally randomize the active power $P$ and reactive power $Q$ by two personalized composite functions $F^p_n(x)$ and $F^q_n(x)$, which is depicted in Fig. \ref{Fig.1}. The constructed composite functions consist of a modified sigmoid function and a constant function, represented by expression \eqref{eq23} - \eqref{eq28}.

$f^p_n(x)$ and $f^q_n(x)$ are the modified sigmoid functions used to introduce nonlinearity into the active and reactive power data transformation, respectively. The parameter $a^p_n$ represents the unique scaling factor for customer $n$ when randomizing active power, while $a^q_n$ is the corresponding scaling factor for reactive power. Both $a^p_n$ and $a^q_n $ are specific to each user.

\begin{align}
    \label{eq23}
    & f^p_n(x) = \frac{1}{1 + e^{-a^p_nx}} - 0.5, \hspace{1.92cm} \forall n \in \mathcal{N}\\
    \label{eq24}
    & g^p_n(x) = c^p_n, \hspace{4.05cm} \forall n \in \mathcal{N}\\
    \label{eq25}
    & F^p_n(x) = f^p_n(x) + g^p_n(x), \hspace{2.205cm} \forall n \in \mathcal{N}\\
    \label{eq26}
    & f^q_n(x) = \frac{1}{1 + e^{-a^q_nx}} - 0.5, \hspace{1.96cm} \forall n \in \mathcal{N}\\
    \label{eq27}
    & g^q_n(x) = c^q_n, \hspace{4.08cm} \forall n \in \mathcal{N}\\
    \label{eq28}
    & F^q_n(x) = f^q_n(x) + g^q_n(x), \hspace{2.28cm} \forall n \in \mathcal{N}
\end{align}

\begin{figure}
    \centering    \includegraphics[width=0.5\textwidth,height=0.20\textwidth]{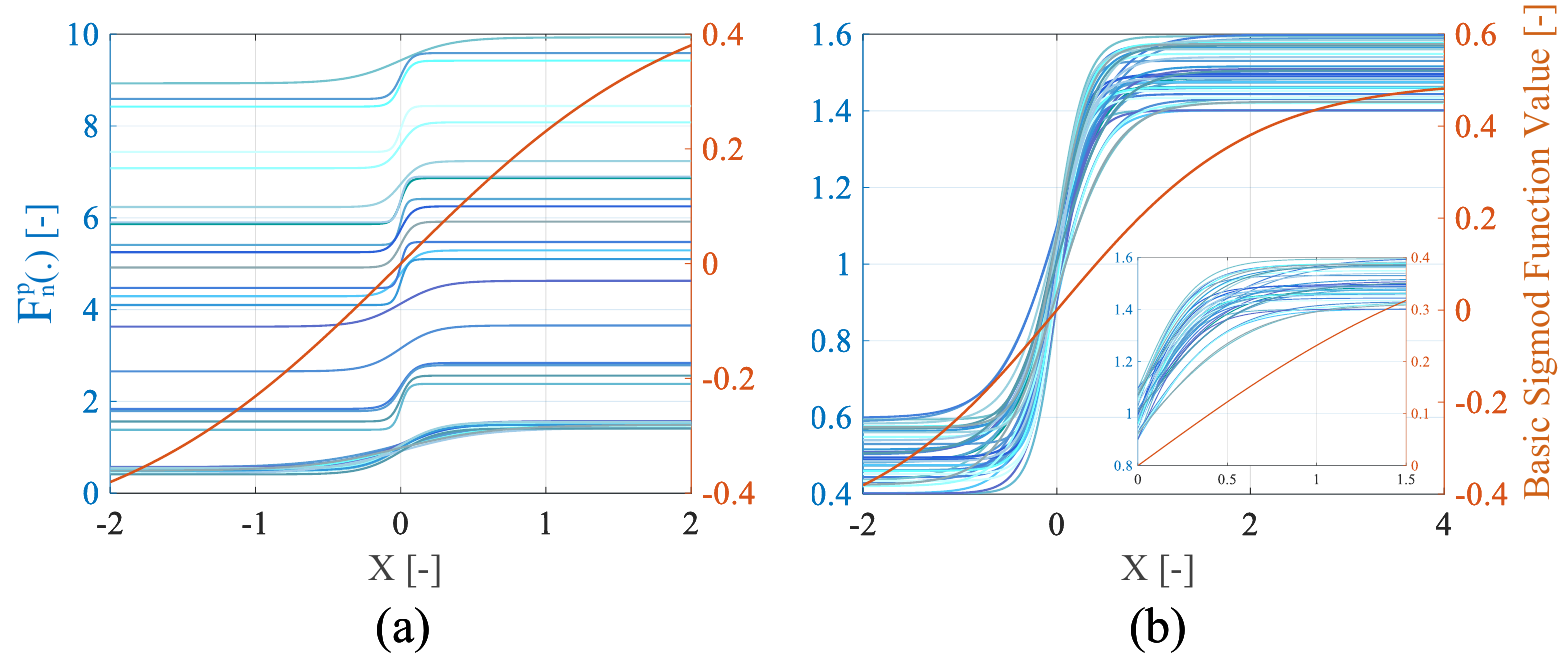}
    \vspace{-0.6cm}
    \caption{Nonlinear transformation function without constraints in (a) and with constraints in (b).}
    \vspace{-0.3cm}
    \label{fig2}
\end{figure}
$ g^p_n(x) $ and $ g^q_n(x) $ represent constant functions, where $ c^p_n $ and $ c^q_n $ are the locally offset noises for active and reactive power, respectively. Similar to the scaling factors, these constant noises are unique for each user and differ between active and reactive power. $ F^p_n(x) $ and $ F^q_n(x) $ are the composite functions combining the logistic transformation and the constant noise for active and reactive power, respectively. This strategy ensures that the randomization transforms the power data according to the unique parameters of each user to weaken the inherent relationship between active and reactive power data. An illustrative example for $F^p_n(x)$ and $F^q_n(x)$ is depicted in Fig. \ref{fig2} (a).

The utility of the transformed power data $P^*$ should be preserved to guarantee the convergence and accuracy of the ANN-based PF calculation. To obfuscate the data (i.e., ensuring the indistinguishability of data from different users) while ensuring data utility, constraints (\ref{eq29}) and (\ref{eq30}) are introduced for parameters $a^q_n$, $a^q_n$, $c^q_n$ and $c^q_n$. An illustrative example of composite functions with constraints is shown in Fig. \ref{fig2} (b).
\begin{align}
\label{eq29}
&\underline{a} \leq a^p_n, a^q_n \leq \overline{a}, \hspace{2cm} \forall n \in \mathcal{N}\\
\label{eq30}
&\underline{c} \leq c^p_n, c^q_n \leq \overline{c}, \hspace{2.2cm} \forall n \in \mathcal{N}
\end{align}
where $\underline{a}, \overline{a},\underline{c}$ and $\overline{c}$ are the upper limits for the parameters in expression (\ref{eq25}) and (\ref{eq28}), respectively. As illustrated in Fig. \ref{fig2} (b), the constrained transformation formula effectively concentrates the data distribution, mitigating potential convergence challenges for the ANN. 

The $\underline{a}, \overline{a},\underline{c}$ and $\overline{c}$ can be determined by assessing the sensitivity of ANN to these parameters and subsequently defining the parameter range. The upper and lower limits of the parameters are also set based on the correlation between the initial data and the transformed data. This approach ensures an optimal balance between data privacy and data utility.

The proposed LRS establishes a robust framework for maintaining data utility while safeguarding sensitive information. Each user (i.e., each SM) in the preset region selects their parameters $a^p_n$, $a^q_n$, $c^p_n$, and $c^p_n$ individually and does not share them with other users and DSO. This individualized selection process significantly enhances data security, as it becomes more challenging to deduce the original data from the encrypted data without knowing the specific parameters used in the randomization process. The received randomized power data are stored $D_L$ at the DSO centre.
\vspace{-0.2CM}
\subsection{ZKP-based Data Collection}
\label{3-2}
\begin{figure}
\centering
\includegraphics[width=0.43\textwidth,height=0.55\textwidth]{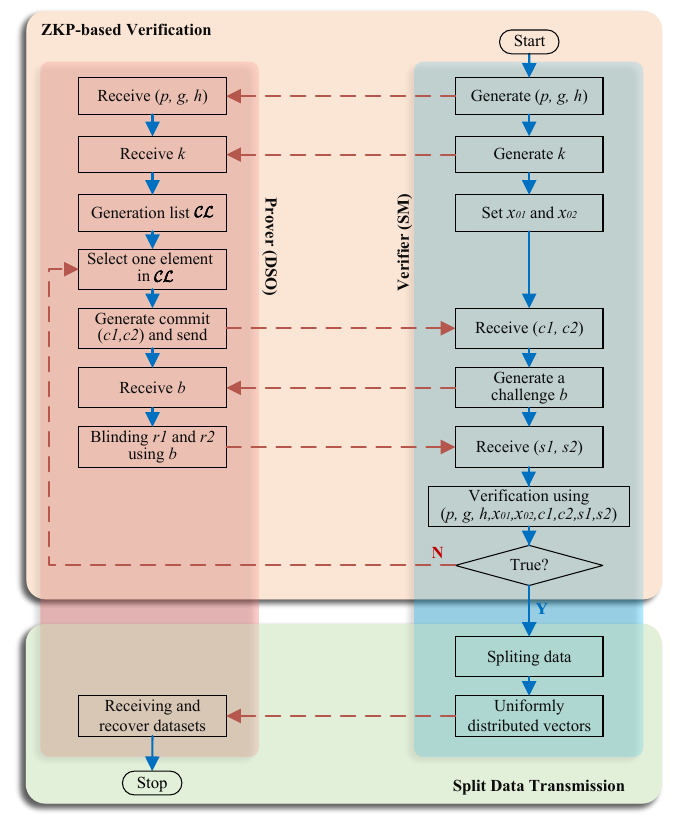}
\vspace{-0.3cm}
\caption{Framework of the ZKP-based data collection process: red lines indicate communication paths and blue lines illustrate the process steps.}
\vspace{-0.5cm}
\label{Fig1.2}
\end{figure}
The framework of the proposed ZKP-based data
collecting strategy is depicted in Fig. \ref{Fig1.2}. In this scheme, the DSO acts as the prover, receiving uniformly distributed vectors $\widetilde{V}$, and proves to the verifier that it knows the index location of the split data. The SM serve as the verifier, sharing only uniformly distributed vectors $\widetilde{V}$ with split voltage magnitudes and verifying the correctness of the DSO's claim.

\begin{figure}
    \centering
    \includegraphics[width=0.28\textwidth,height=0.350\textwidth]{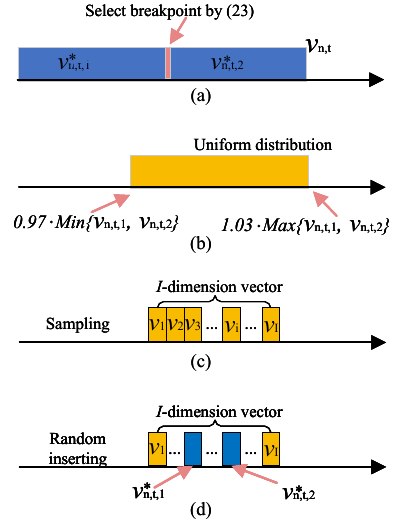}
    \vspace{-0.3cm}
    \caption{Illustrative example of uniformly distributed vectors construction}
    \label{fig4}
    \vspace{-0.5cm}
\end{figure}

The SM randomly splits the voltage magnitudes and inserts them into uniformly distributed vectors $\widetilde{V}$, which are then shared with the DSO. An illustrative example is depicted in Fig. \ref{fig4}. For instance, each voltage magnitude is split into two parts, and vector $\widetilde{V}$ is obtained by \eqref{eq20} - \eqref{eq210}. The upper and lower limits for $\widetilde{V}$ are set as 1.03 $\cdot$ max\{$v^*_{n,t,1}, v^*_{n,t,2}$\} and 0.97 $\cdot$ min\{$v^*_{n,t,1}, v^*_{n,t,2}$\}.
\begin{align}
    \label{eq20}
    &v_{n,t} = v^*_{n,t,1} + v^*_{n,t,2}\\
    &v^*_{n,t,1} = f_r\{0.49\cdot v_{m,n}, 0.51\cdot v_{m,n}\}\\
    \label{eq210}
    &\widetilde{V} = [v_1,\dots ,v^*_{n,t,2},\dots , v^*_{n,t,1}, \dots,v_I]
\end{align}

The sum of any two data points within these vectors $\widetilde{V}$ remains nearly similar, making it difficult to infer the actual voltage values and thereby enhancing privacy. Furthermore, the SM does not provide any information to the DSO on how to extract the original voltage values. Thus, the DSO needs to verify the correct index of each split voltage in the uniformly distributed vectors $\widetilde{V}$ without revealing the true values (i.e., convincing SM that DSO know the true indexes).

As shown in Fig. 3, the SMs first sets the true indexes of the split voltages locally and only shares the dimension $I$ of the uniformly distributed vectors with DSO. Then, the DSO generates a list $\mathcal{CL}$ containing all possible combinations of indices for the true split voltages, with only one being the correct solution. The DSO selects one potential solution and generates two PCs $c1$ and $c2$ using (\ref{eq0}), corresponding to the chosen indices. These PCs are then sent to the SM. Afterwards, the SM sends a challenge $b$ to the DSO, prompting the DSO to compute $s_1$ and $s_2$ using (\ref{eq01}) and the received $b$. The DSO then sends $s_1$ and $s_2$ back to the SM. SM independently decrypts them using \eqref{eq260} and obtain $r_1$ and $r_2$, respectively, 
\begin{equation}
    \label{eq260}
    r = mod\{(s - (1-b) \cdot x_0), p\}
\end{equation}

SM generates the two commitments $c1^*$ and $c2^*$ for the true indexes and compares them with the received commitments $c1$ and $c2$. If they are equal, the SM inform the DSO to receive the time-series dataset and terminates the interaction with this SM. The DSO recovers the true voltage datasets with the true indexes and storage in $D_V$. Each SM will conduct the above process before sharing the split time-series voltage datasets.
\vspace{-0.2CM}
\subsection{ANN-based Power Flow Calculation}
\label{3c}

MLP is employed to conduct ANN-based power flow analysis. The training dataset $D_L$ contains as input data the randonized active power $P$ and reactive power $Q$ collected using the proposed \textit{LRS} strategy in Section \ref{3-1}, while the output dataset contains the voltage magnitudes $D_{V}$ obtained using the ZKP-based collection strategy in Section \ref{3-2}. Consequently, the dimension of the output layer is $N$. The activation function is represented by $\sigma_1$ \dots $\sigma_5$. The loss function is formulated as the expression (\ref{eq21}).
\vspace{-0.3cm}
\begin{equation}   
   \label{eq21}
   \mathcal{L}_{MLP} = \frac{1}{N} \sum_{n=1}^{N} \left[(V_n - \widetilde{V}_n)^2 \right] \quad \quad \quad\quad 
\end{equation}
\begin{equation}
\begin{split}
    \label{eq22}
    &\tilde{V_n} = \sigma_i(W_{i-1} \sigma_{i-1}(...(W_1 \sigma_1(D_n)+ b_1)\\
    & \hspace{3cm} +... + b_{i-1})) +  b_{i} \quad \quad \quad
\end{split}
\end{equation}
where $W_i$ and $b_i$ represent the weights and biases of the hidden layers and will be optimized during the training process. $\tilde{V}_n$ is the estimated voltage magnitude.
\vspace{-0.3CM}

\subsection{Online Application}
\begin{algorithm}[t]
\caption{Model-Free Privacy Preserving PF Analysis in Distribution Networks}
\label{A1}
\KwIn{$D_L$, $D_V$, $D_f$, $l$, $\eta$, $tt^*$,$M_1$}
ANN-2 $\gets$ Train ANN using $D_V$ and $D_L$.\\
$D_f$ $\gets$ $\textit{LRS}$ (P,Q) at each SM\\
Estimate $V$ using ANN-2 and $D_f$\\
\If{$tt \leq tt^*$}{
    $D^*_L$ $\gets$ $\textit{LRS}$ (P,Q) at each SM\\
    $D^*_V$ $\gets$ \textit{ZKP(V)} parallelly\\
    Freeze the last $l$ layers of ANN-2.\\
    \For{$i = 1$ to $M_1$}{
        Train ANN-2 with $\eta$\\
        $\mathcal{L}_{RMLP}$ $\gets$ calculate loss using Eq. (\ref{eq21})\\
        Backpropagate $\mathcal{L}_{RMLP}$\\
        ANN-2 $\gets$ Update ANN-2\\
    }
    return ANN-2\\
}
Return step 2\\
\end{algorithm}
Given datasets $D_L$ and $D_V$, an ANN is trained and stored at the DSO. When the DSO receives transformed data $D_f$ from SMs that employ \textit{LRS}, the trained ANN is utilized for PF analysis. To mitigate the accuracy loss induced by the seasonal variations, the DSO must periodically update the weights of the ANN. Algorithm \ref{A1} provides a detailed description of the complete online application and updating process of the ANN, illustrating how the model evolves to maintain robust performance and accuracy in PF calculations. 

The Wasserstein distance (WD) is employed as an indicator to trigger ANN updates. WD quantifies the cost of transforming one distribution into another, making it a suitable metric for assessing distributional differences. The WD values range from 0 to 1, with 0 indicating high similarity and 1 indicating significant dissimilarity. Specifically, the training voltage datasets are divided into $N_v$ chronologically ordered sub-datasets (e.g., ten-day intervals). The WD $W_i$ between each sub-dataset and the new dataset is calculated. The WD-based indicator, denoted as $tt$, shown in equation (\ref{eq41}).
\begin{equation}
    \label{eq41}
    tt = \sum_{i=1}^{N_v} W_{i} \hspace{0.9cm}
\end{equation}
When the indicator $tt$ surpasses a pre-set threshold, $tt^*$, the ANN updates are triggered. Firstly,  multiple weeks of new SM data are collected, denoted as $D^*_L$ and $D^*_V$. Then, the architecture-based IL strategy, as described in Eq. (\ref{eq18}), is used to update the ANN instead of re-training a new model, i.e., trading off the accuracy and computation cost. The IL strategy requires the specification of a pre-set parameter $l$ and a maximum epoch limit $M1$.

\section{Experimental Results}
\label{case}
In this section, we verify the feasibility and accuracy of the proposed local privacy-preserving strategies on four ANNs and three LVDNs in the Netherlands denoted as LV-52, LV-64, and LV-95 \cite{liu2025topology}. Three datasets are utilized for these LVDNs: 1) time-series profiles for each household with a 15-minute time resolution in LV-52, selected and scaled from \cite{schneider2017analytic}, and 2) time-series profiles with a 1-hour time resolution collected from London, UK and the Netherlands, are used in LV-64 and LV-95 respectively. The power factor is set at 0.95, which is used for PF calculation to obtain voltage datasets.

\subsection{Local Data Randomization}
\begin{figure*}
    \centering
    \includegraphics[width=1.01\textwidth,height=0.50\textwidth]{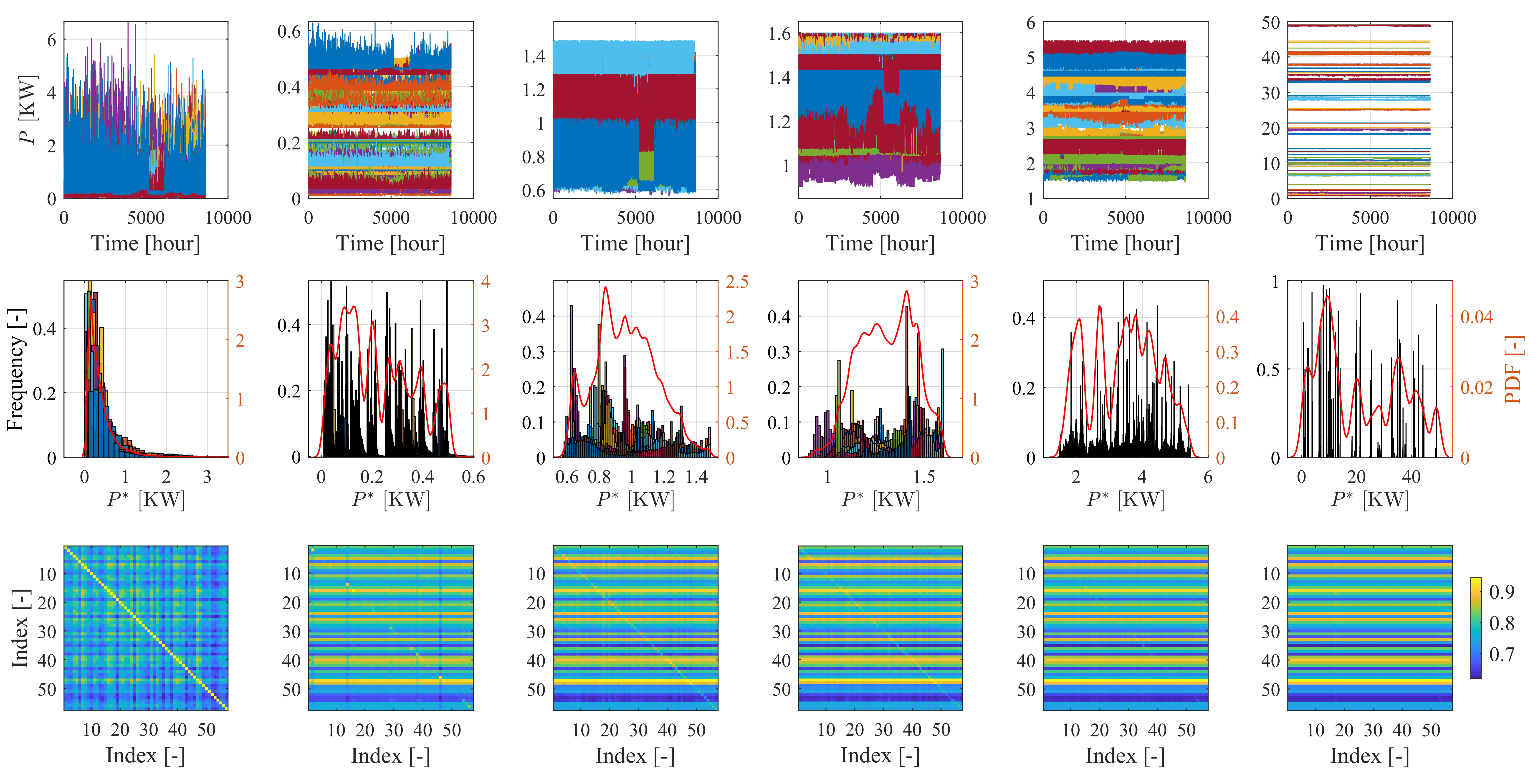}
    \vspace{-0.6cm}
    \caption{Distribution of original $P$ and transformed dataset $P^*$ (in the first and second row) and their correlations (in the third row).}
    \label{fig3}
    \vspace{-0.7cm}
    
    \begin{tabular}{p{0.15\textwidth}p{0.13\textwidth}p{0.14\textwidth}p{0.14\textwidth}p{0.13\textwidth}p{0.15\textwidth}}
        \centering {\scriptsize Set \uppercase\expandafter{\romannumeral1}} & \centering {\scriptsize Set \uppercase\expandafter{\romannumeral2}} & \centering {\scriptsize Set \uppercase\expandafter{\romannumeral3} }& \centering {\scriptsize Set \uppercase\expandafter{\romannumeral4}} & \centering {\scriptsize Set \uppercase\expandafter{\romannumeral5} } & \centering {\scriptsize Set \uppercase\expandafter{\romannumeral6} } \\
    \end{tabular}
    \vspace{-0.5cm}
\end{figure*}

Results on the non-linear transformation of the active power dataset are depicted in Fig. \ref{fig3}, illustrating both the distribution (the second row) and correlation (the third row) before and after the transformation. $P^*$ represents the transformed active power. The first column of results pertains to the original dataset, while the subsequent five columns relate to the transformed data. The upper and lower limits of parameter $a^p_n$ and $a^q_n$ are set as [0, 0.2], [1, 1.5], [3, 8], [8, 10], [0, 50], and for parameter $c^p_n$ and $c^q_n$, the limits are set as [0, 0.5], [0.5, 1], [0.9, 1], [1.2, 5], [0, 50], represented by Set \uppercase\expandafter{\romannumeral1} to \uppercase\expandafter{\romannumeral6}, respectively. The first row displays the data distribution. The second row presents the frequency distribution histogram of each sample and the probability density curve. The distribution of the initial dataset aligns closely with a Beta distribution, which contrasts sharply with the distribution observed in the transformed data. This significant change in distribution enhances the privacy-preserving properties of the data transformation. The third row shows the correlation between the transformed data and the original data, indicating the potential for inferring the original data from the transformed data. When the upper limits of parameters ($a^p_n$, $a^q_n$ $c^p_n$ and $c^q_n$) increases, the correlation between the transformed datasets becomes less distinct, making it challenging to trace the transformed dataset back to the original dataset, as illustrated in columns 3 to 6. However, a larger upper limit also leads to a broader distribution (i.e., in the first row). The expanded distribution range causes the data to be dispersed across different intervals, deviating from the normalized distribution. Thus, the upper limits for parameters $a^p_n$ and $c^q_n$ are set to [3, 8] and [0.9, 1.1], respectively.

\begin{figure}
    \centering
    \includegraphics[width=0.48\textwidth,height=0.21\textwidth]{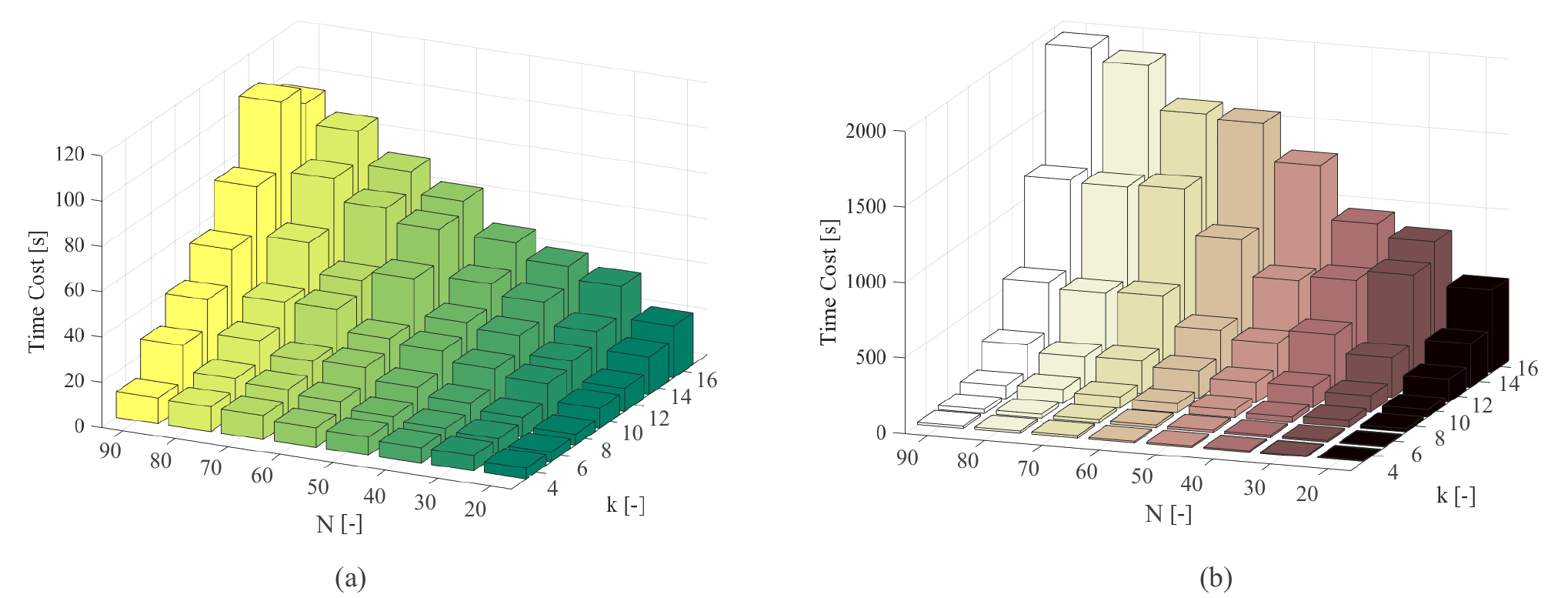}
    \vspace{-0.2cm}
    \caption{Time cost of ZKP-based data collection: (a) with data split into 2 sub-parts, and (b) with data split into 4 sub-parts.}
    \label{fig06}
    \vspace{-0.8cm}
\end{figure}

\subsection{ZKP-based Data Collection Efficiency}
In this case study, we employ the \textit{Secp256k1} elliptic curve for cryptographic operations, which is renowned for its security properties and widespread use in blockchain systems. Communication between the DSO and the SMs is facilitated using ZeroMQ (\textit{zmq} package) with a PAIR socket configuration, ensuring reliable, bi-directional message exchange. All cases are conducted on one laptop with a preset port ID. The impact of parameters $I$ and $N$ on the data collection efficiency is analysed. The maximum values for $I$ and $N$ are set at 16 and 90, respectively. The dimension of the shared dataset is 2800 (i.e., a one-month dataset with 15 min resolution). The time cost is used as an indicator to show the efficiency of the proposed ZKP-based collection strategy, as shown in Fig. \ref{fig06}.

As shown in Fig. \ref{fig06}, the overall time cost increases with both the dimension $I$ of the uniformly distributed vector $\tilde{V}_n$ and the number of SMs sharing data with the DSO. A larger dimension $I$ introduces more potential inserting positions for split data, thereby expanding the search space for the prover (DSO) and increasing the computational burden. Similarly, a higher number of SMs leads to more interactions between the DSO and SMs, further increasing the time cost. Moreover, comparing (a) and (b) in Fig. \ref{fig06}, the time cost induced by the increasing dimension of the uniformly distributed vector $\tilde{V}_n$ is significantly higher than that caused by the increase in the number of SMs. For instance, when $I$ and $N$ are at their maximum values, the time cost for transmitting the uniformly distributed vectors with 2 split data is 97 seconds, while it increases to 1930 seconds for transmitting the uniformly distributed vectors with 4 split data. Nevertheless, the proposed ZKP-based strategy can securely gather data from 90 users in under 40 minutes without compromising data privacy. Besides, in the experiment above, the interactions between the SM and DSO were conducted sequentially. When the interaction is conducted in a parallel environment, the time cost of data collection will further decrease.

\subsection{Accuracy Evaluation}
\begin{figure*}
    \centering
    \vspace{-0.4cm}
    \includegraphics[width=1.02\textwidth,height=0.98\textwidth]{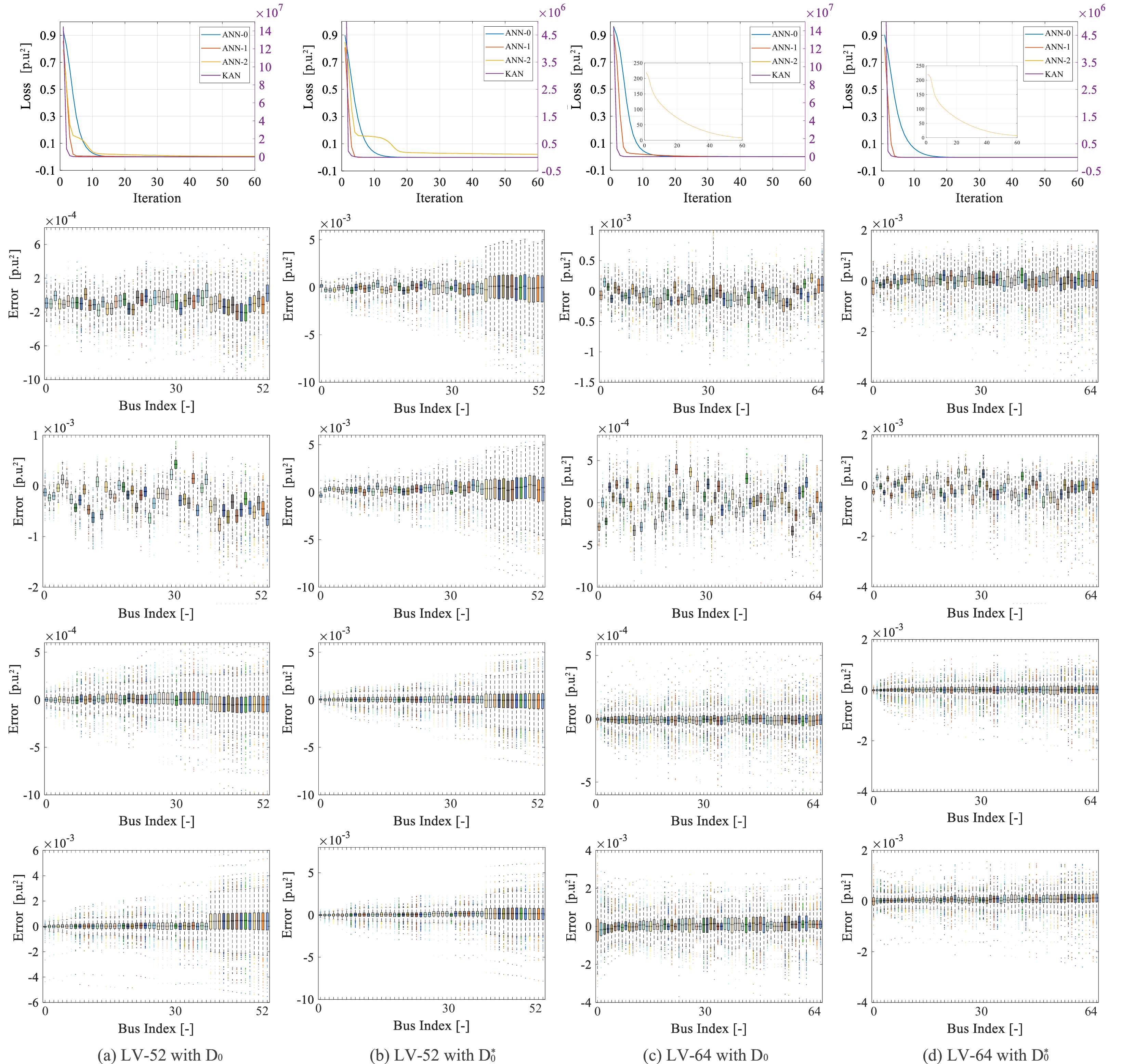}
    \vspace{-0.6cm}
    \caption{The first row presents the convergence curve; the second to fifth rows display the estimation errors for ANN-0, ANN-1, ANN-2, and KAN, respectively.}
    \label{fig5}
    \vspace{-0.5cm}
\end{figure*}
The ANN outlined in \cite{kaseb2023adaptive} is employed as the baseline neural network for PF calculations, referred to as ANN-0. For comparative purposes, a five-layer fully connected ANN and a six-layer  ANN with a scalar are also utilized, designated as ANN-1 and ANN-2, respectively. The learning rate is set to $5.0 \times 10^{-6}$, the maximum number of epochs is 1500, and the batch size is 25. The hidden layer comprises 512 neurons, and the activation function $\tanh$ is used. The AdamW optimizer is employed to train the ANN. Besides, a three-layer  Kolmogorov-Arnold Network (KAN) \cite{liu2024kan}, with an equal number of neurons, is included as a test case. Given the robust capabilities of KAN, the learning rate is set to 0.001, with a smaller batch size of 5, and the maximum number of epochs is set to 150. Both the untransformed dataset $D_0$ and the transformed dataset $D^*_0$ are used to train these four ANNs across the three networks. The input dataset comprises 2,880 samples (i.e., 96 samples per day over 30 days), with 80\% of the data allocated for training and the remaining 20\% for testing. The experiment results are illustrated in Fig. \ref{fig5}.

The first row in Fig. \ref{fig5} demonstrates that all ANNs converge to a value near zero at similar speeds when trained on both the original and transformed datasets. This suggests that the \textit{LRS} does not significantly impact the convergence speed. The four ANNs are trained to the preset maximum number of epochs, and their estimation errors, measured by MSE, are depicted in the second through fifth rows. The results presented in the second and third rows reveal that the error distributions remain within comparable intervals. For instance, the second and fourth columns in these rows indicate errors mainly located in [-0.005,0.005] and [-0.002, 0.002] $p.u.^2$, respectively. This illustrated that increasing the network depth has a negligible impact on improving computational accuracy, highlighting the limitations of this approach in further enhancing performance. Compared to the second and third rows, the mean errors in the fourth row are more concentrated around 0, indicating feature transformation (i.e., the scaler) will sightly mitigate the biase of the estimated voltage magnitude at each node. Furthermore, the fifth row reveals that KAN, despite having fewer layers and fewer training iterations, can achieve accuracy comparable to that of ANN-2 in certain cases, such as with transformed data in the LV-52 and LV-64 networks. Comparing the first with the second columns and the third with the fourth columns, while ANN-0 and ANN-1 perform well on the original data, their accuracy diminishes on the transformed data. In contrast, ANN-2 maintains high accuracy on both the transformed and original datasets, highlighting the effectiveness of the scalar. The performance of the four ANNs under the LV-95 network with the dataset from the Netherlands is similar to their performance under the LV-64 network with the dataset from London (as shown in the third and fourth columns in Fig. \ref{fig5}).

\subsection{Robustness Evaluation}
Considering the random error brought by SMs, three kinds of Gaussian error (i.e., $v_e \sim \mathcal{N}(\mu, \sigma)$) were generated and added to the simulation voltage magnitude data according to the accuracy requirements for SMs \cite{garcia2023phase}. The mean $\mu$ of the Gaussian distribution was set as 0, and three times the standard deviation $3\sigma$ of the Gaussian distribution was set as 0.2\%, 0.5\%, and 1\%. The performance of ANN-2 under LV-64 with transformed data with measurement error is verified. 

\begin{figure}
    \centering
    \includegraphics[width=0.5\textwidth,height=0.25\textwidth]{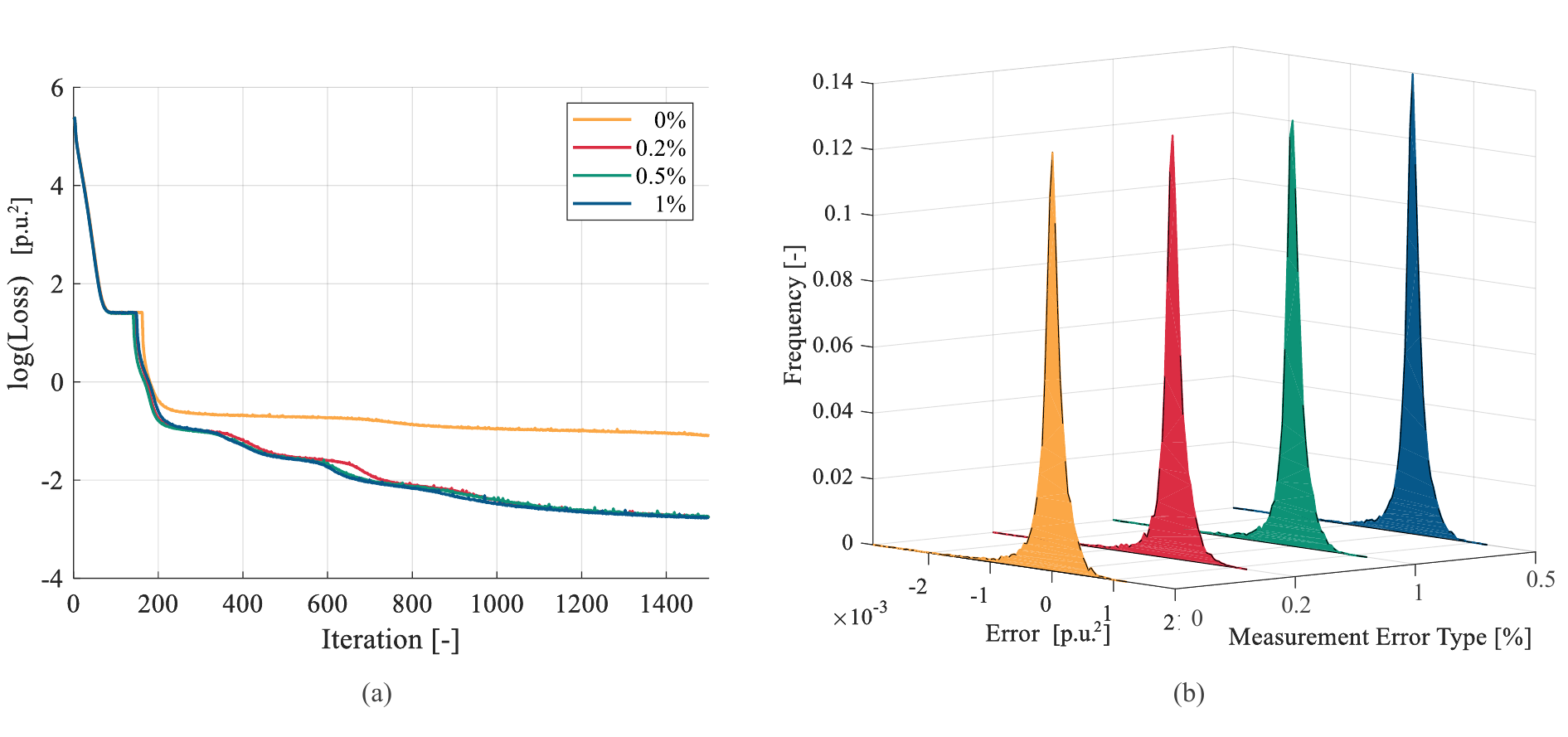}
    \vspace{-0.7cm}
    \caption{Convergence curve of ANN-2 under varying error levels in (a) and probability density distribution of the estimation error in (b).}
    \label{fig6}
    \vspace{-0.3cm}
\end{figure}

The convergence speed of the ANN-2 is depicted in Fig. \ref{fig6} (a). While ANN-2 demonstrates the ability to converge to a lower error value when trained on data with inherent errors, while the convergence process is relatively slow. To assess the robustness of ANN-2, four datasets with varying error levels 0\%, 0.2\%, 0.5\%, and 1\% were employed, with the ANN-2 initially trained on a dataset containing a 0.2\% error. The results, shown in Fig. \ref{fig6} (b), reveal that the error distributions across all test scenarios are strikingly consistent, falling within a narrow range of -0.002 to 0.001 $p.u.^2$. Furthermore, Table \ref{table1} summarizes the testing errors of ANN-2 across 16 different scenarios, which are consistent with the results in Fig. \ref{fig6}. These results indicate that the impact of measurement errors is almost negligible. This underscores the advantage of ANN-based PF calculations over model-based methods: ANN-based approaches exhibit enhanced robustness to errors and inaccurate datasets, leading to more reliable and accurate results.

\vspace{-0.25cm}

\begin{table}
\caption{Test Error ($10^{-4}$) of ANN-2 under Multiple Scenarios}
\label{table1}
\centering
\resizebox{\columnwidth}{!}{
\begin{tabular}{cccccc}
\hline
\hline
\multirow{1}{*}{Test Data} &
\multirow{1}{*}{~} &
\multicolumn {1}{c}{0\%} & 0.2\% &0.5\%&1\% \\
\hline
\multirow {4}{*}{Train Data}
\ {~} & {0\%} & $-0.0697\pm$2.6 & -0.0531$\pm$2.4 & -0.0837$\pm$2.6 & -0.0374$\pm$2.5 \\
\ {~} & {0.2\% } &-0.0945$\pm$2.7 & 0.0466$\pm$2.7 & -0.0576$\pm$2.6 & 0.0166$\pm$2.4 \\
\ {~} & {0.5\%} &  -0.0238$\pm$2.6 & -0.0228$\pm$2.6 & -0.0201$\pm$2.7 & -0.1166$\pm$2.4 \\
\  {~} & {1\%} & -0.0628$\pm$2.6 & -0.1306$\pm$2.5 & -0.0694$\pm$2.6 & 0.0097$\pm$2.9 \\
\hline
\hline
\end{tabular}
}
\vspace{-0.5cm}
\end{table}

\subsection{IL-based Updating}
\begin{figure}
    \centering
    \includegraphics[width=0.5\textwidth,height=0.25\textwidth]{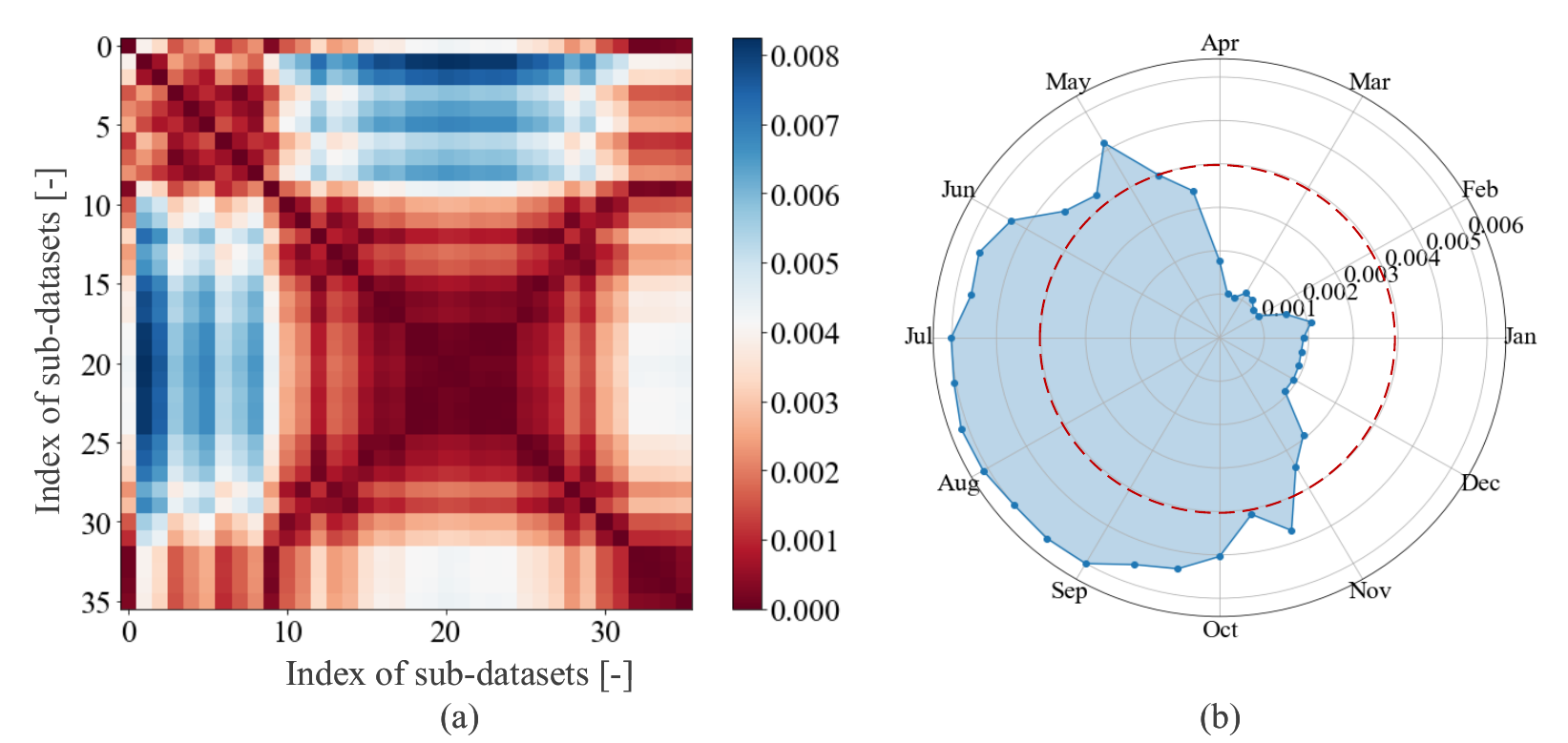}
    \vspace{-0.7cm}
    \caption{WD among 10-day datasets in LV-95 in (a) and the WD-based indicator in (b).}
    \label{fig10}
    \vspace{-0.3cm}
\end{figure}

The one-year voltage dataset from LV-95 was divided into 36 datasets, each covering a 10-day period. The WD between these datasets was calculated and visualized in Fig \ref{fig10} (a). For instance, using the first 10 datasets as the training data, the WD-based indicator $t$ was obtained and displayed in Fig. \ref{fig10} (b). Fig \ref{fig10} reveals significant differences between the distribution of summer and winter datasets, reflecting seasonal variations. The threshold $tt^*$ can be determined by DSO based on historical data or practical experience (e.g., as indicated by the red dashed circle in Fig. \ref{fig10} (b)). 

\begin{figure}
    \centering
    \includegraphics[width=0.5\textwidth,height=0.43\textwidth]{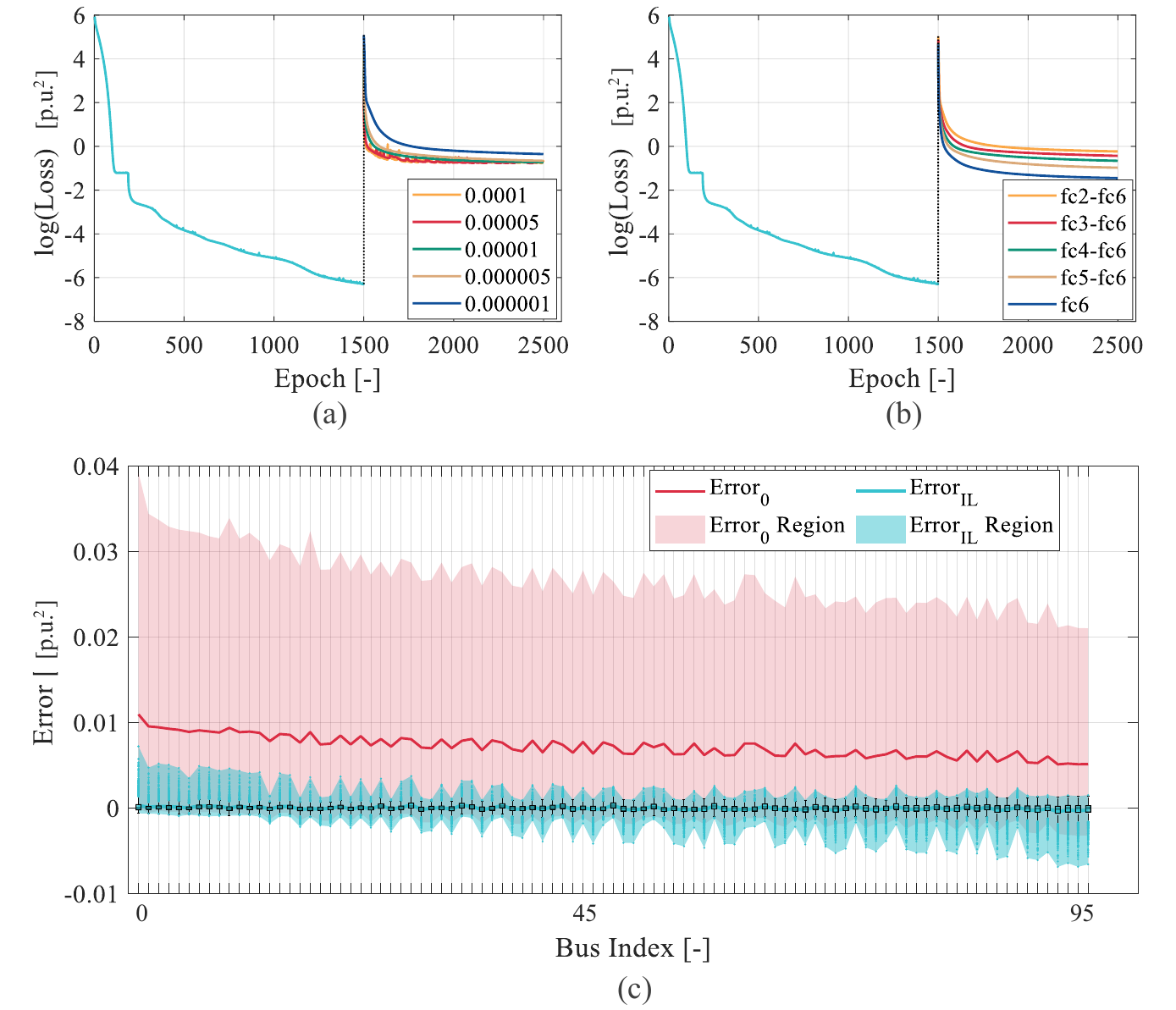}
    \vspace{-0.7cm}
    \caption{Loss curve of ANN-2 with IL under multiple learning rates in (a), multiple frozen layers in (b) and distribution of the estimation error in (c).}
    \label{fig7}
    \vspace{-0.4cm}
\end{figure}

In this paper, the parameters of the last three layers of ANN-2 are frozen as a benchmark, meaning $\boldsymbol{l}$ is set as [4,5,6]. The remaining 3 layers introduced in Section \ref{3c} are updated during the incremental learning period. To assess the feasibility of the employed IL-based updating strategy for ANN-2, a series of experiments were conducted on the LV-95 network across 12 different scenarios, varying the number of frozen layers (ranging from the 2nd to the 6th layer) and adjusting the learning rate (from $5.0 \times 10^{-5}$ to $5.0 \times 10^{-7}$). A three-month dataset in summer (24 samples per day for 90 days) was utilized for initial training. An additional 20-day dataset from winter was employed for IL-based updating, followed by a 60-day dataset used for testing. The maximum number of epochs for updating was set at 1,000. The outcomes of these experiments are detailed in Fig. \ref{fig7} and Table \ref{table2}, which summarize the maximum and mean estimation errors.

Fig. \ref{fig7} demonstrates that the proposed IL-based updating strategy effectively maintains PF accuracy by fine-tuning the parameters of the unfrozen layers. As illustrated in Fig. \ref{fig7} (a), the learning rate has a relatively minor effect on the IL-based updates, with the optimal value for this test case determined to be approximately 0.00001, as indicated in Table \ref{table2}. Compared to the learning rate, the depth of the ANN-2 has a greater influence on the update process and the final convergence value, as shown in Fig. \ref{fig7} (a) and (b), which is consistent with the findings in Table \ref{table2}. Meanwhile, updating more layers (i.e., fewer frozen layers) increases the computational burden. When the learning rate is set to 0.000005 and layers fc4 through fc6 are frozen, the distribution of Error$_{0}$ was depicted in \ref{fig7} (c). The red region in Fig. \ref{fig7} (c) shows a noticeable bias in voltage magnitude estimation when tested on a dataset from a different season, meaning that the trained ANN-2 cannot accurately estimate voltage magnitude for the whole year. However, after applying the IL-based updating strategy, this bias and the region of the estimation error are significantly reduced, as evidenced by the light blue region in Fig. \ref{fig7} (c), indicating the effectiveness of the IL strategy.

\begin{table}
\caption{Test Maximum/Mean Error ($10^{-4}$) of ANN-2 under Multiple Scenarios}
\label{table2}
\centering
\resizebox{\columnwidth}{!}{
\begin{tabular}{ccccc}
\hline
\hline
\multirow{1}{*}{Frozen Layer} &
\multirow{1}{*}{~} &
\multicolumn {1}{c}{fc2-fc6} & fc4-fc6 & fc6 \\
\hline
\multirow {5}{*}{Learning Rate}
\ {~} & {$1\times 10^{-6}$} & 144/0.42 & 89.1/0.04 & 60.2/0.16  \\
\ {~} & {$5\times 10^{-6}$ } & 103/0.05 & 72.1/0.01 & 56.1/0.21  \\
\ {~} & {$1\times 10^{-5}$} &  92.5/0.15 & 71.0/0.09 & 54.9/0.25  \\
\ {~} & {$5\times 10^{-5}$} & 87.3/0.34 & 65.9/0.48 & 66.1/0.53 \\
\ {~} & {$1\times 10^{-4}$} & 83.9/1.97 & 85.0/0.98 & 73.2/1.91 \\
\hline
\hline
\end{tabular}
}
\vspace{-0.6cm}
\end{table}

\section{Conclusion}
\label{conclusion}
This paper introduced a privacy-preserving power flow calculation framework to ensure household privacy in model-free power flow calculation. Compared to differential privacy and traditional encryption strategies, the employed local randomization and ZKP-based data collection strategies are structurally simpler and more easily deployable within SMs. The simulation results indicated that the proposed ZKP-based data collection strategy enables DSO efficient and secure collection of voltage magnitude, and the proposed LRS can effectively change the distribution of the initial data, hindering the true distribution of power datasets. Meanwhile, with the
cooperation of ANN in PF analysis, the constructed ANN maintains a high accuracy on both the original dataset, the randomised dataset and the dataset with measurement errors. Moreover, the results underscore the feasibility of deploying an ANN with an IL-based updating strategy in long-term applications, ensuring sustained accuracy and adaptability.

\bibliography{ref}{}
\bibliographystyle{IEEEtran}

\end{document}